# Real-time Characterization of Gated-Mode Single-Photon Detectors

Thiago Ferreira da Silva, Guilherme B. Xavier, and Jean Pierre von der Weid

*Abstract*—We propose a characterization method for the overall detection efficiency, afterpulse and dark count probabilities of single-photon counting modules in real-time with simple instrumentation. This method can be applied when the module is running in its intended application, and is based on monitoring the statistics of the times between consecutive detections. A mathematical model is derived and fit to the data statistical distribution to simultaneously extract the characterization parameters. The feasibility of our scheme is demonstrated by performing measurements on three commercial devices based on cooled InGaAs APD operating in gated mode. Different statistical ensemble lengths were analyzed and results assess the scheme for real-time application.

## I. INTRODUCTION

Single-photon avalanche photodiodes (SPADs) have become a key element in ultra-sensitive applications, such as quantum communication [1], photon counting optical time-domain reflectometry [2], optical metrology with higher resolution than the classical limit [3], sensor applications [4] and imaging for biomedical purposes [5]. The characterization of SPADs is an important benchmark when employing them into practical applications. In general, there are three physical performance parameters of a SPAD that are of interest. They are: the dark count rate (false detections generated from thermal excitation, tunneling electric carriers, etc.); the detection efficiency (overall detection efficiency considering the external optical coupling losses on the photodiode, the electric carriers generation by absorbed photons and the avalanche triggering by the photo-generated carriers) and the afterpulse probability (the probability of some trapped carriers created during a previous avalanche being released and inducing a subsequent avalanche, originating a false detection). Many specialized characterization methods have been developed over the years for SPADs [6-9].

For a long time the detection of single-photons had been mainly based on photo-multiplier tubes [10]. More recently avalanche photodiodes have become a very interesting alternative, as they are more practical to use [11,12], can be more easily integrated into systems, and can be deployed in arrays [13]. They can be typically divided into two types, depending on the operation mode: free-run and gated. For free-running detectors, passive or active quenching circuits must be provided to both reset the device after a detection event and protect it from damage due to the macroscopic current flow [12]. Regardless of their type they have the same basic principle of operation: they are reverse-biased above the breakdown region, such that even a single photon can originate an avalanche of electric carriers by the cascaded effect of the impact ionization. The macroscopic current is monitored and formatted into a voltage pulse, which represents a detection event [12]. SPADs working in passive free-run mode are constantly placed just above of the breakdown region. After an avalanche is generated, a voltage drop due to a large resistor in series with the diode removes it momentarily from the breakdown region, and allows it to recover for the next detection. In active mode, the avalanche is sensed by an electronic circuit, which then actively removes the diode from the breakdown region, allowing for a faster recovery than passive quenched SPADs [11,12]. This free-run active mode of operation is typically used in Si SPADs [12,15].

For use in long-distance quantum communication, Si SPADs are not suitable since they do not absorb photons in the 1550 nm telecom spectral region, ideal for fiber-optical propagation [15]. For this purpose InGaAs-based SPADs have been developed and used in many long-distance experiments [16,17]. Due to their inherent higher noise, when compared to Si-based SPADs, they are usually operated in the so-called gated mode [11,12,15]. In this mode the detector is reverse biased just out of the breakdown region. When the single-photon is expected to arrive, a short voltage pulse is combined with the reverse bias applied to the detector, such that diode is

This work was supported by Brazilian agencies CNPq, CAPES and FAPERJ. Additionally G. B. Xavier would like to acknowledge financial support from Milenio project NC10-030-F.

T. Ferreira da Silva is with the Optical Metrology Division, National Institute of Metrology, Standardization and Industrial Quality, Duque de Caxias, RJ 25250-020 Brazil (phone: +55 21 2679 9051; fax: +55 21 2679 9207; e-mail: tfsilva@inmetro.gov.br) and also with the Center for Telecommunications Studies, Pontifical Catholic University of Rio de Janeiro.

G. B. Xavier is with the Departamento de Ingeniería Eléctrica, Universidad de Concepción, Casilla 160-C, Concepción, Chile and also with the Center for Optics and Photonics, Universidad de Concepción, Casilla 4016, Concepción, Chile (email: guilherme.xavier@cefop.udec.cl).

J. P. von de Weid is with the Center for Telecommunication Studies, Pontifical Catholic University of Rio de Janeiro, Rio de Janeiro, RJ 22451-900 Brazil (e-mail: vdweid@opto.cetuc.puc-rio.br).

kept above the breakdown region for the duration of the pulse. This time period is generally called the gate or detection window, and typical values are in the order of a few ns.

Recent progress in gating technology aims on extending the gate frequency capability of SPAD's while limiting the afterpulse effect. The techniques rely basically on quenching the avalanches as soon as possible, which is achieved by making the detector able to sense small avalanches. In the self-differencing technique, the electrical [18] or optical [19] detection signal is subtracted from a delayed replica of the gate response. This allows the reduction of the strong signature imposed on the detection signal by the finite capacitance of the APD. As weak avalanches can be sensed, they can be quickly quenched and the afterpulse effect is reduced. Another method employs a sine wave to gate the detector and a filter to mitigate the spurious signal [20]. Hybrid techniques have also been related, combining features of both methods [21]. These technologies make SPAD's suitable for operation under gigahertz gating frequencies [22-24] and allowed demonstration of photon-number resolution capability [25] which further extends the range of applications of APD-based single-photon detectors.

In this paper we propose a characterization method for gated-mode SPADs, which yields the performance parameters of the device in real-time, allowing its monitoring under real operating conditions using simple instrumentation. We first explain in Sec. II the characterization method, based on the acquisition of the times between consecutive detections of the SPAD, which are organized in a histogram, and describe the mathematical model used to extract the parameters from the histogram. In Sec. III we describe the experimental setup used to validate our method and we focus the experimental work on commercial InGaAs SPADs (id201, idQuantique) working in gated mode. In Sec, IV we show the results from the experimental characterization of three SPADs, varying different settings such as, repetition rate, gate width, overall detection efficiency and average photon number per gate window. We compare the results obtained with values provided by the manufacturer and also verify how fast the method is capable of providing a reliable measurement of the SPAD. Finally the obtained results and conclusions of the method are provided in Sec. V.

## II. CHARACTERIZATION METHOD

The characterization method is based on the analysis of the statistical distribution of times between consecutive detections from the SPAD [9]. As we focus on gated mode operation, we will assume that detection events can occur only during a gate window. Furthermore, in this case we can define the measured times as multiples $m$ of the gating period $T$. Given that the SPAD generated an initial detection, the time elapsed until the occurrence of the next detection event is recorded, independently of what caused it (whether it is a count triggered by an arriving photon, an afterpulse, or a dark count). This is depicted in Fig. 1, with $m = 2$ gates elapsed between the first and second detections for example.

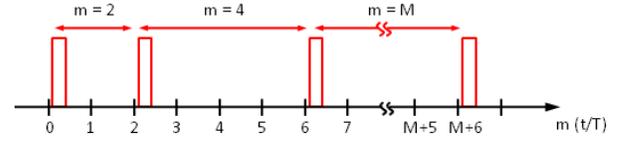

Fig. 1. Principle of the acquisition of time intervals.

The system will then record the number of gates (which is equivalent to the time elapsed) between the second detection and the third one, and so on for the adjacent detections, as shown in Fig. 1. This procedure is repeated continuously until a sufficient statistical ensemble is collected. The recorded times are grouped together in a histogram, which, after normalization, shows the probability distribution function of detection times for the SPAD under the current operational condition.

According to the figure, the random variable $m$ assumes the value $M$ whenever $M-1$ empty gates are followed by a detection event, which can be written as:

$$P(m) = [1 - P_{nc}(m)] \times \prod_{x=1}^{m-1} P_{nc}(x) \quad (1)$$

where $P_{nc}(m)$ is the probability of no counts occurring at the $m^{th}$ time slot. No count is registered if neither a dark count, nor an afterpulse nor the detection of a photon takes place. Equation (1) is rewritten substituting the dark count probability per gate ($P_d$) and the afterpulse probability at the $m^{th}$ gate ($P_a(m)$), as well as the probability of having no incident photons detected in a gate ($P_{n=0}$), resulting in:

$$P(m) = \{1 - P_{n=0} \cdot (1 - P_d) \cdot [1 - P_a(m)]\} \times \\ [P_{n=0} \cdot (1 - P_d)]^{m-1} \times \prod_{x=1}^{m-1} [1 - P_a(x)] \quad (2)$$

Assuming that the photon number distribution of an attenuated laser source in a given time window is Poissonian and considering the overall detection efficiency ($\eta$) of the device, we have $P_{n=0} = exp(-\mu\eta)$, where $\mu$ is the average number of photons per gate [26]. By rewriting Eq. (2) with other distributions, other types of photon sources with different photon number statistics could be applied to the model. The afterpulse probability as a function of the number of gates elapsed is modeled assuming a dominant single exponential decay for the trapped carriers with detrapping lifetime $\tau$, resulting in $P_a(m) = P_0 exp(-mT/\tau)$, where $P_0$ is an amplitude related to the number of filled traps, which is treated as a fitting parameter.

By expanding the product of terms at the end of Eq. (2) and applying the summation of finite geometric progressions, it becomes

$$\prod_{x=1}^{m-1}[1-P_a(x)]=1-\alpha+\beta, \quad (3)$$

where the terms $\alpha$ and $\beta$ correspond to the first and second order expansions respectively, and are given by:

$$\alpha = P_0 e^{-T/\tau}\left[e^{-(m-1)T/\tau}-1\right]/\left(e^{-T/\tau}-1\right) \quad (4)$$

$$\beta = \frac{P_0^2}{e^{-T/\tau}-1} \times \left\{\frac{e^{-(m+1)T/\tau}\left[e^{-(m-2)T/\tau}-1\right]}{e^{-T/\tau}-1} - \frac{e^{-3T/\tau}\left[e^{-(m-2)2T/\tau}-1\right]}{e^{-2T/\tau}-1}\right\} \quad (5)$$

The third order term was calculated but we verified that it does not significantly affect the results. The final equation of the model to fit the data is then given by

$$P(m) = \{1-\exp(-\mu\eta)\cdot(1-P_d)\cdot[1-P_0\exp(-mT/\tau)]\} \times [\exp(-\mu\eta)\cdot(1-P_d)]^{m-1}\times(1-\alpha+\beta) \quad (6)$$

The parameters $P_d$, $P_0$, $\tau$ and the product $\mu\eta$ are obtained by fitting Eq. (6) to the acquired data.

Finally the total afterpulse probability ($P_T$) can be obtained by summing the afterpulse probabilities $P_a(m)$ over time from the first gate to infinity, resulting in:

$$P_T = P_0 \frac{1}{e^{T/\tau}-1}, \quad (7)$$

From the model we see that the values of $\mu$ and $\eta$ are coupled. As a consequence, to extract one of them it is necessary to assume we have prior knowledge of the other one.

### III. EXPERIMENTAL SETUP

In order to perform the experiment we used a continuous-wave (cw) laser diode, whose output wavelength is in the telecom window (1550 nm), shown in Fig. 2 as DL.

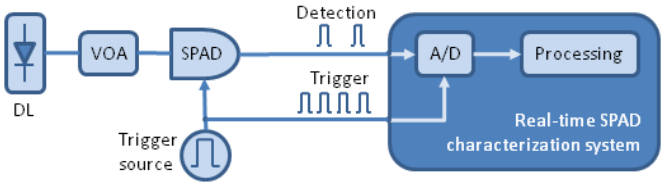

Fig. 2. Experimental setup.

The fiber pigtailed output of the laser is connected to a variable optical attenuator (VOA), from which we can set the $\mu$ average photons per detection window. The output of the attenuator is then connected to the SPAD and an external signal generator is used to provide the trigger pulses to the detector. Both the trigger signal and the electrical output from the SPAD under characterization are connected to the acquisition system, composed of a fast A/D board and a personal computer. Three commercial SPADs (named #1, #2 and #3) are characterized using the proposed method. For each one we set, in the device, the nominal gate width and overall detection efficiency. For each measurement run, with a set of operational condition parameters (average photon number per detection window and gating period), data was continuously acquired by the acquisition system, which sampled both the trigger pulses from the signal generator and the electrical output of the SPAD. The data was processed in custom-made software to extract the number of gate pulses elapsed between consecutive count events and these values are grouped in a normalized histogram in logarithmic scale. The probability density function is then automatically and in real-time fit with the logarithm of the model in Eqs. (4-6), using the number of gates instead of time.

### IV. RESULTS AND DISCUSSIONS

Each of the three SPADs was measured under several combinations of gate frequencies and average number of photons per gate. By fitting the model to the distribution of acquired time intervals between consecutive counts, the overall detection efficiency, dark counts and afterpulse probabilities were extracted. The value of $\mu$ is known and adjusted with a power meter and a variable optical attenuator. An auxiliary measurement consists of the characterization of the effective gate time window. This was performed with the aid of a heralded single-photon source [27]. This is based on the generation of pairs of single-photons through the process of spontaneous parametric down-conversion in a non-linear crystal. When one of the photons of a pair is detected in a trigger detector, it provides an electrical signal which indicates the temporal localization of its twin. The electrical trigger signal passes through a variable delay generator and triggers the SPCM at the desired time, shifted relative to the emitted probe photon. The normalized count rate obtained for a delay scan of the trigger signal is used to measure the response of the SPAD inside the gate. The effective temporal gate width is then calculated, which directly influences on the correct value of $\mu$. Despite the fact that we have used a heralded single-photon source, a laser producing short pulses synchronized to detector gates and an electrical delay generator are sufficient for this task. The procedure and results are similar to our measurements, with an optical pulse being temporally scanned relative to the detection window to obtain its profile.

The histograms with the statistical distribution of times between consecutives detections of SPAD #2, collected at different gate frequencies (200, 400, 600, 800 and 1000 kHz) with a fixed value of 0.08 photons per gate, are grouped in Fig. 3a. In Fig. 3b, data was taken with SPAD #3 at a gate frequency of 600 kHz for different values of $\mu$ (0.04, 0.08, 0.16 and 0.32). In each measurement series, a total of 2 x 10$^6$ detections was acquired. The fit of our model to the data sets is exhibited, with the first 15 gates highlighted in the figure insets.

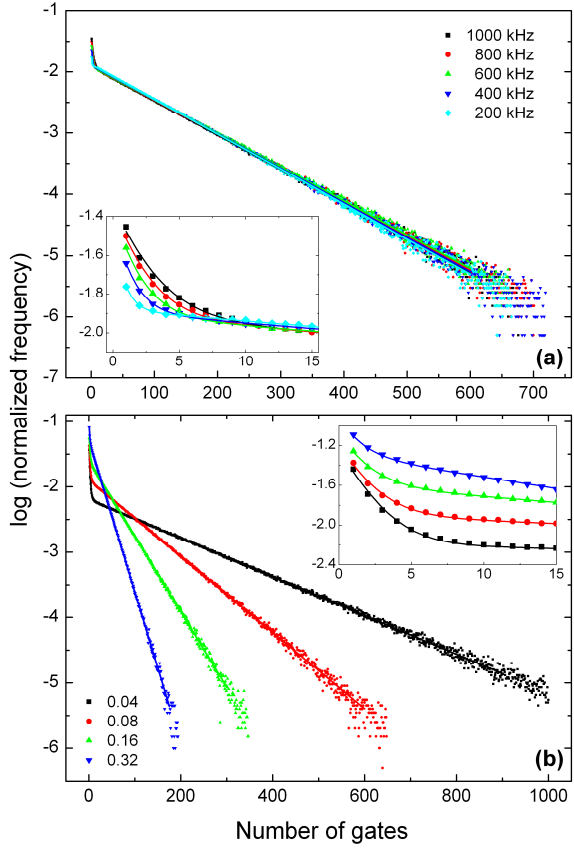

Fig. 3. Model fit to experimental data with different (a) gate frequencies (SPAD #2) and (b) average number of photons (SPAD #3). Insets: zoom in on the first gates.

The histograms present two regions with distinct behaviors. The smooth straight line, corresponding to a decaying exponential in the linear scale, is related to the time independent sources of avalanches, i.e., the counts originated by detected photons and dark counts. Due to the definition of the mathematical model, each point in the histogram depends on the non-occurrence of a count at every gate before it. In other words, when a given time interval of $M$ gates is recorded, this means that there occurred $M-1$ gates without the generation of an avalanche. As the occurrence of longer time intervals is conditioned to having more gates without counts of any kind, this type of event is less frequent. Therefore the slope is expected to be negatively inclined. Observe in Fig. 3a that the histograms are presented as a function of the number of gates, so the slopes are the same for each gate frequency, as the average number of photons per detection window is constant. It is worth noting that, multiplying the number of gates by the gate period, this representation will change and the curves will no longer be overlapped. In this case, for a higher gate frequency, the slope of this curve gets more inclined. The average photon number dependence can be observed in Fig. 3b. As this value increases, there is a higher probability that at least one photon is encountered in a gated time window, as can be verified in the Poissonian distribution. This makes the events closer to each other and the slope of the curve increases (in this case, the $\eta$ parameter is constant).

The sharp region on the beginning of each histogram, highlighted on the figure insets, corresponds to the afterpulse probability. The exponentially decaying behavior is dependent on the gate frequency and the higher this is, the higher the contribution of the afterpulses on the total events probability. The total afterpulse probability for the three detectors was characterized with the same nominal gate width (5 ns) and detection probability (15 %) for different gate frequency values. The parameter was averaged for each gate frequency over four measurements with $\mu$ set at 0.04, 0.08, 0.16 and 0.32 photons per effective gate, as the afterpulse probability is not expected to significantly change with $\mu$. Table 1 summarizes the average afterpulse probability (in percentage) for the three SPADs and the respective standard deviation values. All three detectors present the same qualitative behavior concerning the total afterpulse probability evolution according to the change in gate frequency.

TABLE I
TOTAL AFTERPULSE PROBABILITY AT DIFFERENT GATE FREQUENCIES FOR EACH SPAD

| Gate frequency (kHz) | #1 | #2 | #3 |
|---|---|---|---|
| 200 | 0.56 ± 0.03[1] | 1.12 ± 0.05 | 0.67 ± 0.03 |
| 400 | 1.79 ± 0.05 | 3.62 ± 0.08 | 2.10 ± 0.03 |
| 600 | 3.28 ± 0.08 | 6.47 ± 0.08 | 3.81 ± 0.08 |
| 800 | 4.83 ± 0.22 | 9.98 ± 0.31 | 5.72 ± 0.15 |
| 1000 | 6.47 ± 0.19 | 13.52 ± 0.36 | 7.71 ± 0.27 |

[1] Values in percentage: average ± standard deviation.

The values obtained for the overall detection efficiency and dark count probability of the devices were averaged over all measurement series performed for each combination of gate frequency and $\mu$ (see legends of Fig. 3 for values), totalizing 20 series of $2 \times 10^6$ points for each detector. The devices were set to 15 % and 5 ns nominal overall detection efficiency and gate width, respectively. The final values are summarized in Table II.

TABLE II
$\eta$ AND $P_D$ MEASURED FOR EACH SPAD

| SPAD | $\eta$ (%) | $P_d$ (×10$^{-4}$) |
|---|---|---|
| #1 | 15.10 ± 0.15[1] | 1.6 ± 0.8 |
| #2 | 15.07 ± 0.09 | 2.1 ± 0.9 |
| #3 | 15.03 ± 0.03 | 1.8 ± 0.8 |

[1] Average ± standard deviation.

The mean values for overall detection efficiency are very close to the nominal ones. The values for the dark count probability were alternatively characterized by triggering the detectors at low gate frequency (10 kHz) with the optical input covered. This frequency was chosen to avoid significant contribution from afterpulses. The values were averaged over 30 measurements with integration times of 20 s for each detector, yielding the probability ($\times 10^{-4}$) of 1.1, 1.5 and 1.2, respectively for #1, #2 and #3, with standard deviation of $0.5\times 10^{-4}$, in agreement with the previous results. All the results for the afterpulse probability characterization were compared to the values obtained by fitting a straight line (in the logarithmic scale) to the histogram, without the initial points (the bend). Assuming that all points above the line are due to afterpulses, the ratio between the area under the fit straight line and the area under the whole set of data points was computed, as seen in Fig. 4.

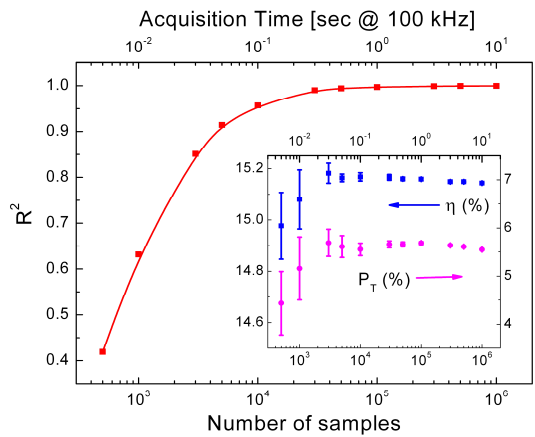

Fig. 5. Goodness of fit to statistical datasets with different lengths (line is a guide to the eye). Inset: extracted parameters. Error bars represent experimental standard deviation of the mean.

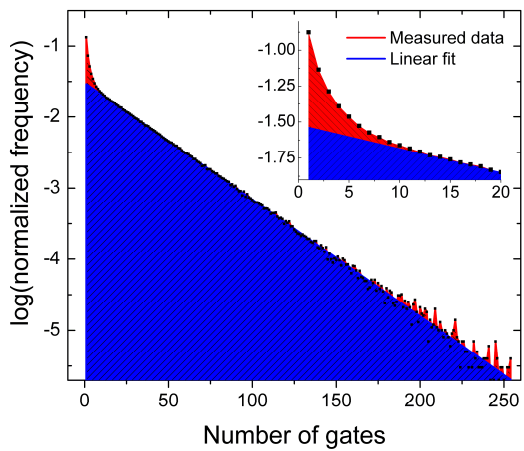

Fig. 4. Analysis of the total afterpulse probability by the areas ratio.

The average ratio between these values and the afterpulse probabilities extracted with the proposed model agreed, for the detectors under all combination sets of gate frequencies and $\mu$, within 2.2 %, indicating that in our case the single-trap model is sufficient. An extension to a multiple-trap case is also possible.

Figure 5 shows the $R^2$ parameter for the fit of our model to the acquired data according to the number of samples taken. This value quantifies the goodness of the fit and is expected to be as close as possible to unity.

Data corresponds to the average of the results over 20 consecutive runs of the real-time characterization system, with different number of samples for each run. As expected, the increase in sample statistics makes the fit more representative, with the $R^2$ value better than 0.99 with $3\times 10^4$ points and better than 0.999 with $3\times 10^5$. The inset presents the average and standard deviation of the parameters $\eta$ and $P_T$ extracted from the fits. As the length of the samples decreases, the statistical spread increases, but the average values still show good agreement with the expected values. Even for $3\times 10^3$ samples, the relative experimental standard deviation of the mean results in 0.26 % for the overall detection efficiency and 5.0 % for total afterpulse probability. These results assess our method as suitable for real-time monitoring of the single-photon detector under operational conditions.

## V. Conclusions

We presented a method for characterization of SPADs by analyzing the times between consecutive detection events. Real-time monitoring of the device working under operational conditions was achieved with simple instrumentation. The measurements were performed with the detector working in a standard configuration, enabling the overall detection efficiency, dark counts and afterpulse probabilities to be simultaneously extracted by application of the deduced analytical model. Conversely, if the SPAD overall detection efficiency is known, the incident average number of photons can be extracted. Results from the characterization of three commercial devices based on gated mode APDs were presented and results agreed with auxiliary measurements. An analysis of the characterization results for different statistical sample lengths demonstrated the feasibility of our method for real-time monitoring of the detector under operational mode.

### Acknowledgment

The author would like to acknowledge Guilherme P. Temporão for helpful discussions.